\documentstyle[aps,multicol]{revtex}

\renewcommand{\narrowtext}{\begin{multicols}{2} \global\columnwidth20.5pc}
\renewcommand{\widetext}{\end{multicols} \global\columnwidth42.5pc}
\begin{document}

\title{Optimal quantum pumps have a vanishing transmission coefficient}

\author{Anton Alekseev}

\address{
University of Geneva, Section of Mathematics, 2-4 rue du Li\`evre,
c.p. 240, 1211 Gen\`eve 24, Switzerland \\
Uppsala University, Institute for Theoretical Physics,
Box 803, S-75108, Uppsala, Sweden}

\date{January 2002}

\maketitle

{\tightenlines
\begin{abstract}

\hskip -0.3cm
PACS numbers: 73.23.-b, 72.10.-d

\end{abstract}
}

\narrowtext

In a recent Letter, Avron {\em et. al} \cite{Avron} introduced a notion
of {\em optimal quantum pumps}. These are adiabatic 
quantum pumps which work without dissipation. In particular,
they produce neither  entropy nor noise. 
In the present Comment we show that in the absence of  magnetic field
optimal quantum pumps always have a vanishing
transmission coefficient. Such `quantum pumps'
do not make use of Quantum Mechanics since all tunneling
or interference effects are banned by vanishing of the
transmission coefficient. 
We leave it as an outstanding question whether
genuine optimal quantum pumps with nonvanishing transmission
coefficient can be constructed by making use
of the magnetic field.

Recall that a  quantum pump is a mesoscopic device attached to two
(or more) reservoirs by means of ideal quantum wires.
All reservoirs have the same chemical potential $\mu$,
and when the pump is not at work no charge transfer
occurs between reservoirs. 
In the scattering approach 
pioneered by B\"uttiker {\em et al.}
\cite{But1} and developed in \cite{Br,But2} a quantum pump
connected to $n$ reservoirs is described by the scattering
matrix at the Fermi energy $S_{kl}$ where the indices $k$ and $l$
label the outgoing and incoming channels, respectively. 
The matrix $S_{kl}$ is always unitary, 
$S^{-1}_{kl}=S^*_{lk}$. As the pump
operates, the scattering matrix changes with period $\tau$,
$S(t+\tau) = S(t)$. In practice, this is usually achieved 
by applying an alternating voltage to some gates inside
the mesoscopic device \cite{exp}.
The net current pumped into the
channel $k$ at the moment $t$ is given by formula
(eqn. (3) of \cite{Avron}),
\begin{equation} \label{current}
J_k = \frac{ie}{2\pi} \, (\partial_t S S^\dagger)_{kk}=
\frac{ie}{2\pi} \, \sum_{l=1}^n (\partial_t S_{kl}) S^\dagger_{lk}
\end{equation}
For the noise production one gets 
(eqn. (10) of \cite{Avron}),
\begin{equation} \label{noise}
N_k = \frac{\beta h e^2}{24\pi^2} \sum_{l \neq k} 
| (\partial_t S S^\dagger)_{kl} |^2.
\end{equation}
 
In the absence of 
magentic field, ${\bf B}= \nabla \times {\bf A} =0$,
one can always choose a gauge in which the vector
potential ${\bf A}$ vanishes. Then, 
the Schr\"odinger equation which 
governs the motion of electrons through the pump
has {\em real coefficients} since
all the terms which contain $i$ are proportional
to ${\bf A}$.
This implies that
taking a complex conjugate of a solution yields again a
solution. Choosing the incoming and outgoing wave functions
complex conjugate to each other (this corresponds to
turning $e^{ikx}$ to $e^{-ikx}$) one obtains 
an extra symmetry of the scattering matrix
$S^{-1}_{kl}=S^*_{kl}$. Together with unitarity
this implies that the scattering matrix is symmetric.

For simplicity, we consider quantum pumps with two external
reservoirs. In this case, the general form of a unitary
symmetric scattering matrix is given by formula,
$$
S = 
\left(
\begin{array}{cc}  
a  &  b \\
b  &  c
\end{array}
\right) \, ,
$$
where $b$ is the transmission amplitude, and 
$a$ and $c= - a^*b/b^*$ are reflection amplitudes.
The matrix $\partial_t S S^\dagger$ is anti-Hemritian,
$$
\partial_t S S^\dagger =  
\left(
\begin{array}{cc}  
i x_1  &  y \\
- y^*  &  i x_2
\end{array}
\right) \, .
$$
The requirement that $\partial_t S$ is symmetric yields
$yc + y^*a=i(x_2-x_1) b$. Formula (\ref{noise}) now reads,
\begin{equation} \label{key}
N_{1,2}=\frac{\beta h e^2}{24 \pi^2} |y|^2 \ge
\frac{\beta h}{6} \, \frac{T}{1-T} \, J^2,
\end{equation}
where $T= |b|^2$ is the transmission coefficient
and $J=(J_1-J_2)/2=e(x_2-x_1)/4\pi$ is the net
current between the reservoirs. 
We conclude that in the absence of magnetic
field an optimal quantum pump ($N_{1,2}=0$) is only
possible if $T=0$. A surprising denominator $(1-T)$
in formula (\ref{key}) accounts for the fact that
a transparent quantum pump with $T=1$ cannot
transfer charge between reservoirs, and 
in this case $J=0$.

{\bf Acknowledgements}. I am grateful to M. B\"uttiker,
J.-P. Eckmann and G.-M. Graf for the discussion which
initiated this work and for their valuable comments.
I greatly appreciate remarks by V. Cheianov
and L. Sadun.

\widetext

\end{document}